\documentclass[aip,amsmath,amssymb,reprint]{revtex4-1}
\usepackage{balance}
\usepackage{mathptmx}
\usepackage{amsmath}
\usepackage{amssymb}
\usepackage{graphicx} 
\usepackage{float}
\usepackage{fancyhdr}
\usepackage[english]{babel}
\addto{\captionsenglish}{%
  
}
\usepackage{array}
\usepackage[T1]{fontenc}
\usepackage[usenames,dvipsnames]{xcolor}
\usepackage{setspace}
\usepackage[compact]{titlesec}
\usepackage{hyperref}

\usepackage{color}
\usepackage{soul}

\makeatletter
\def\@email#1#2{%
 \endgroup
 \patchcmd{\titleblock@produce}
  {\frontmatter@RRAPformat}
  {\frontmatter@RRAPformat{\produce@RRAP{*#1\href{mailto:#2}{#2}}}\frontmatter@RRAPformat}
  {}{}
}%
\makeatother
\begin{document}
\preprint{AIP/123-QED}
\title[]{Electrical noise in electrolytes: a theoretical perspective}
\author{Th\^e Hoang Ngoc Minh}
\thanks{These authors contributed equally to this work}
\affiliation{
Sorbonne Universit\'e, CNRS, Physicochimie des \'Electrolytes et Nanosyst\`emes Interfaciaux, F-75005 Paris, France
}
\author{Jeongmin Kim}
\thanks{These authors contributed equally to this work}
\affiliation{
Sorbonne Universit\'e, CNRS, Physicochimie des \'Electrolytes et Nanosyst\`emes Interfaciaux, F-75005 Paris, France
}
\author{Giovanni Pireddu}
\affiliation{
Sorbonne Universit\'e, CNRS, Physicochimie des \'Electrolytes et Nanosyst\`emes Interfaciaux, F-75005 Paris, France
}
\author{Iurii Chubak}
\affiliation{
Sorbonne Universit\'e, CNRS, Physicochimie des \'Electrolytes et Nanosyst\`emes Interfaciaux, F-75005 Paris, France
}
\author{Swetha Nair}
\affiliation{
Sorbonne Universit\'e, CNRS, Physicochimie des \'Electrolytes et Nanosyst\`emes Interfaciaux, F-75005 Paris, France
}
\author{Benjamin Rotenberg}
\thanks{Email: benjamin.rotenberg@sorbonne-universite.fr}
\affiliation{
Sorbonne Universit\'e, CNRS, Physicochimie des \'Electrolytes et Nanosyst\`emes Interfaciaux, F-75005 Paris, France
}
\affiliation{R\'eseau sur le Stockage Electrochimique de l'Energie (RS2E), FR CNRS 3459, 80039 Amiens Cedex, France}

\date{\today}

\begin{abstract}
Seemingly unrelated experiments such as electrolyte transport through nanotubes, nano-scale electrochemistry, NMR relaxometry and Surface Force Balance measurements, all probe electrical fluctuations: of the electric current, the charge and polarization, the field gradient (for quadrupolar nuclei) and the coupled mass/charge densities. The fluctuations of such various observables arise from the same underlying microscopic dynamics of the ions and solvent molecules. In principle, the relevant length and time scales of these dynamics are encoded in the dynamic structure factors. However modelling the latter for frequencies and wavevectors spanning many orders of magnitude remains a great challenge to interpret the experiments in terms of physical process such as solvation dynamics, diffusion, electrostatic and hydrodynamic interactions between ions, interactions with solid surfaces, etc. Here, we highlight the central role of the charge-charge dynamic structure factor in the fluctuations of electrical observables in electrolytes and offer a unifying perspective over a variety of complementary experiments. We further analyze this quantity in the special case of an aqueous NaCl electrolyte, using simulations with explicit ions and an explicit or implicit solvent. We discuss the ability of the standard Poisson-Nernst-Planck theory to capture the simulation results, and how the predictions can be improved. We finally discuss the contributions of ions and water to the total charge fluctuations. This work illustrates an ongoing effort towards a comprehensive understanding of electrical fluctuations in bulk and confined electrolytes, in order to enable experimentalists to decipher the microscopic properties encoded in the measured electrical noise. 
\end{abstract}

\maketitle


One of the most important properties of electrolytes, consisting of ions in a solvent, and more generally of ionic fluids, is their ability to conduct electricity under an external electric field. The resulting electric current results from the interplay between the driving force, interactions between ions, and interactions between the ions and the solvent molecules -- generally understood in terms of electrostatic and hydrodynamic effects, as well as diffusion (thermal fluctuations) and friction (dissipation). Since ions and solvent molecules display a (multipolar) charge distribution, their coupling with electromagnetic fields is at the heart of many experimental techniques probing a wide frequency range, including dielectric relaxation spectroscopy~\cite{kremer_broadband_2003, buchner_dielectric_2004}, infrared and Raman THz spectroscopy~\cite{balos_time-resolved_2022}, which can fruitfully be combined with measurements of the conductivity at low frequency~\cite{balos_macroscopic_2020}. The interpretation of these experiments in terms of microscopic mechanisms, in particular solvation dynamics, greatly benefits from molecular simulations~\cite{maroncelli_computer_1988, raineri_molecular_1994, jimenez_femtosecond_1994, stirnemann_mechanisms_2013, laage_effect_2019, buchner_interactions_2009, heyden_dissecting_2010, popov_mechanism_2016, mamatkulov_water-separated_2018, banerjee_ions_2019, carlson_exploring_2020}, or analytical theories for the frequency-dependent conductivity~\cite{chandra_frequency_1993, chandra_frequency_2000, yamaguchi_theoretical_2007}.

The analysis of the fluctuations of the electric current, via its power spectral density, revealed an algebraic behaviour at low frequency (``$1/f$'', or ``coloured noise'') in bulk electrolytes~\cite{hooge_1_f_1970, vasilescu_electrical_1974} as well as, more recently, in experiments involving ionic currents through single nanopores~\cite{hoogerheide_probing_surface_2009, heerema_1_f_2015, secchi_scaling_2016}. These observations prompted a number of theoretical and simulation studies to assess its microscopic origin~\cite{zorkot_current_2016, zorkot_power_2016, zorkot_current_sPNP_2018, mahdisoltani_long_2021, peraud_fluctuating_2017, gravelle_adsorption_2019, marbach_intrinsic_2021}. Such electrical noise is also exploited in electrochemical impedance measurements~\cite{bertocci_noise_1995, wang_electrochemical_2021, vivier_impedance_2022} as well as in nanofluidic setups using electrodes~\cite{zevenbergen_electrochemical_2009, mathwig_electrical_2012}. The charge fluctuations of electrodes can also be analyzed to investigate the interfacial properties of nanocapacitors in simulations~\cite{limmer_charge_2013, scalfi_charge_2020, scalfi_molecular_2021, cats_capacitance_2021}. As discussed in more detail in Section~\ref{sec:Fluctuations}, fluctuations of the electrostatic potential or the electric field experienced by an atom, which are intimately related to the dynamics of its microscopic environment, in particular the solvent polarization, plays an essential role on electron transfer reactions~\cite{marcus1956a, marcus1965a}, water autodissociation~\cite{geissler_autoionization_2001,hassanali_recombination_2011} as well as ion pair dissociation~\cite{geissler1999a, ballard_toward_2012, kattirtzi2017a}. The fluctuations of the electric field gradient (EFG) drive the nuclear magnetic resonance (NMR) relaxation of quadrupolar nuclei, so that these fluctuations also provide in principle information on the microscopic fluctuations around the latter~\cite{abragam1961principles}. Here again, molecular simulations prove very useful to quantitatively model the EFG fluctuations and open the way to the interpretation of quadrupolar NMR relaxation in terms of molecular motion~\cite{engstrom_molecular_1984, badu_quadrupolar_2013, carof_accurate_2014, carof_microscopic_2015, carof_collective_2016, philips_quadrupolar_2017, philips_quadrupolar_2020, mohammadi_nuclear_2020, chubak_nmr_2021, gimbal-zofka_simulations_2022, chubak_quadrupolar_2023}.

The dynamics of charge fluctuations are also related to the static correlations between ions, as well as with the polar solvent. These correlations are generally understood in terms of screening: of the electrostatic interactions between ions by the solvent (with the reduction of the Coulomb interaction by the permittivity of the latter) and of the electrostatic potential by the ions (with the canonical Debye-H\"uckel theory and corresponding screening length). The issue of static correlations between ions and solvents has regained interest in recent years due to the report of long-range forces in Surface Force Balance experiments with ionic liquids and concentrated electrolytes~\cite{gebbie_ionic_2013, gebbie_long-range_2015, lee_scaling_2017, lee_underscreening_2017}, with an ``anomalous underscreening'' at odds with the Debye-H\"uckel picture. From the dynamical point of view, linear response theory provides a practical route to determine the frequency-dependent conductivity or permittivity from simulations using Green-Kubo or Einstein-Helfand relations involving the appropriate correlation functions of the electric current or polarization, even though the separation between mobile charges and polar molecules is a subtle issue~\cite{sega_dielectric_2015, sega_calculation_2013, cox_finite_2019}. Confining electrolytes, or even pure solvent, between neutral, charged or metallic walls, introduces further complexity, as this modifies the static and dynamic correlations between the polar molecules and the ions. This changes, sometimes dramatically when the distance between the confining walls decreases below tens of nanometers, the static and frequency-dependent permittivity~\cite{ballenegger_local_2003, ballenegger_dielectric_2005, rotenberg_frequency-dependent_2005, gekle_anisotropy_2012, schlaich_water_2016, fumagalli_anomalously_2018,  loche_breakdown_2018, loche_universal_2020, santos_consistent_2020, mondal_anomalous_2021, olivieri_confined_2021, cox_dielectric_2022, underwood_dielectric_2022}, or the spectroscopic response~\cite{ruiz-barragan_nanoconfinement_2022}. 

The spatial and temporal correlations of the charge density are quantified by the charge-charge intermediate scattering function or the charge-charge dynamic structure factor. The former is a function of wavenumber $k$ and time $t$, whose initial value is the static structure factor, while the latter is a function of $k$ and frequency $\omega$. They can be determined and analyzed in molecular simulations (see \emph{e.g.} Refs.~\citenum{bopp_frequency_1998, omelyan_longitudinal_1998, ladanyi_computer_1999} for pure water). The wavenumber- and frequency-dependent response of the current and polarization to electric field is directly related to the corresponding conductivity and permittivity tensors~\cite{fulton_dipole_1978, giaquinta_collective_1978, felderhof_fluctuation_1980, pollock_frequency-dependent_1981, caillol_theoretical_1986, caillol_dielectric_1987}. While these quantities cannot be measured directly, most of the observables corresponding to the experiments described above can be expressed as special cases ($k\to0$ for the macroscopic limit, $t\to0$ or $\omega\to0$ for the static limit) or as weighted integrals over modes. Therefore, these experiments provide, at least in principle, complementary windows to observe the charge fluctuations over different spatial and temporal scales.

Here, we highlight the central role of the charge-charge dynamic structure factor in the fluctuations of electrical observables in electrolytes and offer a unifying perspective over seemingly unrelated experiments. We further analyze this quantity in the special case of an aqueous NaCl electrolyte, using simulations with explicit ions and an explicit or implicit solvent. We discuss the ability of the standard mean-field Poisson-Nernst-Planck theory to capture the simulation results, and how the predictions can be improved. We finally discuss the contributions of ions and water to the total charge fluctuations. Section~\ref{sec:Fluctuations} provides an overview of electrical fluctuations in electrolyte, introducing the relevant quantities and their link with various experimental observables. Section~\ref{sec:Szz} then presents the theoretical and numerical approaches used in this work. Finally, the results are reported and discussed in    Section~\ref{sec:Results}.

\section{Electrical fluctuations in electrolytes}
\label{sec:Fluctuations}
 
\subsection{Charge density and electric current fluctuations}
\label{sec:Fluctuations:DensityCurrent}

We consider the dynamics of an ensemble of $N$ classical particles with (partial) charges $q_i=z_i e$, where $e$ is the elementary charge and $z_i$ the valency. The microscopic state of the system if characterized by their positions $\vec{r}_i(t)$ and velocities $\vec{v}_i(t)$, from which one can determine the instantaneous charge density
\begin{align}
    \rho_q(\vec{r},t) &= \sum_{i=1}^N q_i \delta\left[ \vec{r}_i(t) - \vec{r} \right]
    \label{eq:RhoZofr}
\end{align}
and the electric currrent density
\begin{align}
    \vec{j}_q(\vec{r},t) &= \sum_{i=1}^N q_i \vec{v}_i(t)\delta\left[ \vec{r}_i(t) - \vec{r} \right]
    \, ,
    \label{eq:JZofr}
\end{align}
where the $q$ subscript refers to the electric charge and $\delta$ denotes the Dirac delta function. The fluctuations of these quantities are conveniently analyzed in reciprocal and frequency space, so that we introduce the following spatial Fourier transform
\begin{align}
    \hat{A}(\vec{k}) &= \int_V A(\vec{r}) e^{-i\vec{k}\cdot\vec{r}} \,{\rm d}\vec{r}
    \, ,
    \label{eq:FourierSpace}
\end{align}
where $V$ is the volume of the system, and temporal Laplace transform (with Laplace variable $s=-i\omega$)
\begin{align}
    \tilde{B}(\omega) &= \int_0^{\infty} B(t) e^{+i\omega t} \,{\rm d}t
     \, .
     \label{eq:LaplaceTime}
\end{align}
For the charge density, this leads to
\begin{align}
    \hat{\rho}_q(\vec{k},t) &= \sum_{i=1}^N q_i e^{-i\vec{k}\cdot\vec{r}_i(t)}
        \label{eq:RhoZofk}
\end{align}
from which we can define the charge-charge intermediate scattering function~\cite{hansen_mcdonald_theory_of_simple_liquids_book}
\begin{align}
    F_{qq}(\vec{k},t) &= \frac{1}{N}\langle\hat{\rho}_q(\vec{k},t)\hat{\rho}_q(-\vec{k},0)\rangle
    \, ,
    \label{eq:DefFzz}
\end{align}
where $\langle \cdots \rangle$ denotes an ensemble average.
Other normalizations by the volume instead of the number can be found in the literature. For a bulk isotropic system, this quantity depends only on the norm $k$ of the wavevector. The initial value (for $t=0$) of $F_{qq}$ is the charge-charge static structure factor
\begin{align}
    S_{qq}(\vec{k}) &= F_{qq}(\vec{k},t=0) = \frac{1}{N}\langle \left|\hat{\rho}_q(\vec{k},t=0)\right|^2\rangle
    \, ,
    \label{eq:DefSzzStatic}    
\end{align}
while the Laplace transform of $F_{qq}$ provides the charge-charge dynamic structure factor
\begin{align}
    S_{qq}(\vec{k},\omega) &= \int_{-\infty}^{\infty} F_{qq}(\vec{k},t) e^{+i\omega t} \,{\rm d}t = \tilde{F}_{qq}(\vec{k},\omega) + \tilde{F}_{qq}(\vec{k},-\omega) 
    \, .
    \label{eq:DefSzzofkandomega} 
\end{align}
$F_{qq}(\vec{k},t)$ can be recovered by the inverse Fourier transform
\begin{align}
    F_{qq}(\vec{k},t) &= \frac{1}{2\pi} \int_{-\infty}^{\infty}  S_{qq}(\vec{k},\omega) e^{ -i\omega t} \,{\rm d}\omega \, .
\end{align}
The above quantities cannot all be measured directly in experiments as a function of wavenumber $k$ and frequency $\omega$, even though they are related in various ways to a number of experimental properties, in particular the response of the system to an external electric field. Some examples of observables will be introduced in Section~\ref{sec:Fluctuations:Observables}. Similar functions can be defined from other densities, weighed \emph{e.g.} by the mass or the neutron scattering lengths of each atom instead of their charge, and the corresponding scattering functions are related to responses other than the charge or current induced by an electric field, which will not be considered here (see \emph{e.g.} Ref.~\cite{sedlmeier_charge/mass_2014} for electro-acoustic couplings in pure water). Nevertheless, we emphasize that combining the responses to various perturbations provides complementary windows on the dynamics of the particles -- just as \emph{e.g.} X-ray and neutron diffraction provide complementary information on the structure of water~\cite{soper_joint_2007, amann-winkel_x-ray_2016}.

\subsection{Link with various observables}
\label{sec:Fluctuations:Observables}

Having introduced the quantities describing the charge fluctuations in Section~\ref{sec:Fluctuations:DensityCurrent}, we now discuss the link between the latter and various properties. We begin by standard ones related to the current or polarization response of a bulk liquid to an external field in Section~\ref{sec:Fluctuations:Observables:CurrentPolarization}. We then introduce less frequently considered observables such as the electric field gradient in Section~\ref{sec:Fluctuations:Observables:PotentialFieldEFG}, or the charge induced by an electrolyte on a metallic electrode in Section~\ref{sec:Fluctuations:Observables:Electrode}.

\subsubsection{Electric current and polarization}
\label{sec:Fluctuations:Observables:CurrentPolarization}

The response of the charge distribution to external electric fields is usually investigated by introducing monochromatic perturbations $\vec{E}^{ext}(\vec{k},\omega)$. One is then interested in the electric current  $\vec{j}_q(\vec{k},\omega)$ (defined from Eq.~\ref{eq:JZofr} using Eqs.~\ref{eq:FourierSpace} and~\ref{eq:LaplaceTime}). It is common practice to distinguish ``free'' and ``bound'' charges, by separating the net charge from higher moments of each molecule, in particular their dipole moment, and to introduce the polarization density $\vec{P}(\Vec{r},t)$, whose divergence is minus the density of bound charges. The total electric current is then separated into a term for free charges and a polarization current (time derivative of the polarization) and the effect of an external field on the current $\vec{j}_q(\vec{k},\omega)$ and polarization $\vec{P}(\vec{k},\omega)$ are described with response functions, which are related to the wavenumber and frequency-dependent permittivity and conductivity. These quantities are defined by introducing the Maxwell field $\vec{E}(\vec{k},\omega)$ inside the sample, which differs from the external one due to the screening of the latter by the system itself. The conductivity tensor is defined by Ohm's law (here with Fourier instead of Laplace transforms):
\begin{align}
    \vec{j}_q(\vec{k},\omega) &= \mathbf{\sigma}(\vec{k},\omega) \cdot \vec{E}(\vec{k},\omega) 
    \label{eq:DefSigma}
\end{align}
and can be further separated into longitudinal and transverse conductivities
\begin{align}
    \mathbf{\sigma}(\vec{k},\omega) &= \frac{\vec{k}\,\vec{k}}{k^2} \sigma_l(\vec{k},\omega) + \left[ \mathbf{I} - \frac{\vec{k}\,\vec{k}}{k^2} \right] \sigma_t(\vec{k},\omega)
    \, ,
    \label{eq:DefSigmaLandT}
\end{align}
with $\mathbf{I}$ the identity tensor. The permittivity tensor, defined by
\begin{align}
    \vec{P}(\vec{k},\omega) &= \epsilon_0\left[ \mathbf{\epsilon}(\vec{k},\omega) -\mathbf{I}\right]\cdot \vec{E}(\vec{k},\omega) 
    \, ,
    \label{eq:DefEpsilon}
\end{align}
can similarly be split into longitudinal and transverse components. 

Such a separation between free and bound charges is however not necessary in principle, and experiments such as dielectric spectroscopy probe in fact both contributions simultaneously~\cite{fulton_dipole_1978, caillol_theoretical_1986, schroder_computation_2008, sega_calculation_2013, kremer_broadband_2003}. The relations between conductivity or permittivity and the microscopic response, described below, can be formulated using the total electric current, or equivalently the total polarization, which includes the contribution of ``free'' charges via the so-called itinerant polarization (the time-integral of the corresponding current)~\cite{sega_calculation_2013}. The total charge and polarization are related by $\nabla\cdot\vec{P}=-\rho_q$, and the total current and polarization are then related by $\vec{j}_q=\partial_t\vec{P}$ (for a more complete description at the continuum level, see Ref.~\citenum{sprik_electric-field-based_2021}). This last relation, together with Eqs.~\ref{eq:DefSigma} and~\ref{eq:DefEpsilon} lead to the relation between the generalized conductivity and permittivity~\cite{sega_calculation_2013,hansen_mcdonald_theory_of_simple_liquids_book}
\begin{align}
     \mathbf{\sigma}(k,\omega) &= -i\omega \, \epsilon_0 \left[ \mathbf{\epsilon}(k,\omega) - \mathbf{I} \right]
     \, ,
     \label{eq:LinkSigmaEpsilon}
\end{align}
which is usually used in dielectric spectroscopy for $k=0$. In this context, one sometimes also introduces the apparent permittivity (see \emph{e.g.} Ref~\citenum{sega_calculation_2013}) $\epsilon(\omega) + i\sigma(0)/\epsilon_0\omega$, such that in the limit $k\rightarrow 0$ and $\omega \rightarrow 0$  Eg.~\ref{eq:LinkSigmaEpsilon} reduces to the static conductivity, which can be computed as a Green-Kubo integral of the current autocorrelation function~\cite{hansen_mcdonald_theory_of_simple_liquids_book}. 

For their part, the response functions, or susceptibilities, express the change in the charge, polarization or current induced by the external field $\vec{E}^{ext}(\vec{k},\omega)$. Their link with the permittivity and conductivity depends on the boundary conditions (in particular in simulations using periodic boundary conditions)~\cite{ felderhof_fluctuation_1980, pollock_frequency-dependent_1981,caillol_dielectric_1987} and on whether one considers the longitudinal or transverse response. When retardation effects can be neglected, for $k\neq0$ the response of the polarization is related to the permittivity by~\cite{caillol_dielectric_1987,ladanyi_computer_1999}:
\begin{align}
\chi_l(k,\omega) &= 1-\frac{1}{\epsilon_l(k,\omega)}
\, ,
\label{eq:ChiL}
\\
\chi_t(k,\omega) &= \epsilon_t(k,\omega)-1
\label{eq:ChiR}
\; .
\end{align}

Linear response theory then provides the expression of the relevant response functions, in terms of equilibrium fluctuations of the electric current and polarization~\cite{caillol_dielectric_1987}. For example, for non-polarizable systems the above correlation function is related to $S_{qq}(\vec{k})$ and $\tilde{F}_{qq}(\vec{k},\omega)$ as~\cite{giaquinta_collective_1978, madden_consistent_1984, ladanyi_computer_1999, hansen_mcdonald_theory_of_simple_liquids_book}
\begin{align}
    \chi_l(\vec{k},\omega) &= \frac{\beta N}{V \epsilon_0 k^2 } \left[ S_{qq}(\vec{k})+ i\omega \tilde{F}_{qq}(\vec{k},\omega) \right]
    \, ,
    \label{eq:Chi}
\end{align}
where $\beta=1/k_BT$, with $k_B$ Boltzmann's constant and $T$ the temperature. Since $S_{qq}(\vec{k},\omega)=2\operatorname{Re}\left[\tilde{F}_{qq}(\vec{k},\omega)\right]$ (from Eq.~\ref{eq:DefSzzofkandomega} and the fact that $F_{qq}(\vec{k},t)$ is an even function of time), one also finds this result as a relation between $\operatorname{Im}\left[\chi_l(\vec{k},\omega)\right]$ and $S_{qq}(\vec{k},\omega)$, and the real part is obtained from the Kramers-Kronig relation. Note that this form of fluctuation-dissipation relation holds in the classical limit, which restricts in principle the range of frequencies to $\omega\ll k_BT/\hslash$, with $\hslash$ the reduced Planck constant~\cite{bopp_frequency_1998}. Some molecular vibrations may correspond to frequencies not satisfying this constraint, but in the application to aqueous electrolytes we will use a rigid water model for molecular simulations, which suppresses high frequency vibrations. As mentioned above, the present discussion is also limited to classical nonpolarizable models of the charge distribution, and we note that \emph{ab initio} descriptions may lead to further complications (see \emph{e.g.} Ref.~\citenum{grasselli_topological_2019}). 

The $k^2$ factor in the denominator of the r.h.s. of Eq.~\ref{eq:Chi} reflects the link between polarization and charge, which reads in Fourier space $\tilde{\rho}_q=i\vec{k}\cdot\vec{P}$. Using the initial value theorem for Laplace transforms, $\lim_{\omega\to\infty} \left[ -i \omega \tilde{F}_{qq}(k,\omega) \right] = F_{qq}(k,t=0) = S_{qq}(\vec{k})$ so that $\lim_{\omega\to\infty}\chi_l(\vec{k},\omega)=0$ and $\lim_{\omega\to\infty}\epsilon_l(\vec{k},\omega)=1$. As mentioned above, in dielectric spectroscopy experiments one only has access to the $k\to0$ limit, so that the response functions are not known directly as a function of $k$ and $\omega$. However, different experiments can provide complementary information on $F_{qq}(k,t)$ (or other quantities related to the dynamics of charge fluctuations), as illustrated \emph{e.g.} in Section~\ref{sec:Fluctuations:Observables:PotentialFieldEFG}.

While we have mainly emphasized the linear response of the electric current to electric fields, we note that previous works have also considered electrokinetic couplings from the cross-correlations of electric and mass currents~\cite{marry_equilibrium_2003, sedlmeier_charge/mass_2014, yoshida_molecular_2014, mangaud_sampling_2020}. Following early theoretical studies on coupled transport phenomena in ionic fluids~\cite{giaquinta_collective_1978}, it might also be possible to extract transport coefficients from various density fluctuations, as recently proposed for heat conductivity in uncharged systems~\cite{cheng_computing_2020}. In addition, recent methodologies based on the large deviations of the electric current fluctuations have also been introduced to predict the response to large external fields, including the couplings between ions and solvent~\cite{lesnicki_field_2020, lesnicki_molecular_2021}.

\subsubsection{Electric potential, field and field gradient}
\label{sec:Fluctuations:Observables:PotentialFieldEFG}

Fluctuating sources (charges) result in fluctuating electric potential $\phi$ (scalar), field $\vec{E} = -\nabla \phi$ (vector) and field gradient $\mathbf{V} = -\nabla \nabla\phi$ (rank-2 tensor). The solvation dynamics around ions can be probed in spectroscopic experiments and some observables are related to the change in the electronic distribution between the ground state and excited states, which couples to the electric potential, field or field gradient fluctuations~\cite{maroncelli_computer_1988, raineri_molecular_1994, jimenez_femtosecond_1994}. These fluctuations and their mutual effect on ions or complex solutions can be sampled in molecular simulations~\cite{reischl_statistics_2009, matyushov_nanosecond_2011, martin_non-gaussian_2012, sellner_charge_2013, samanta_ionic_2022} or modelled by analytical approaches based on continuum electrostatics~\cite{hynes_continuum_1981, stenhammar_electric_2009, stenhammar_unified_2011}, (Gaussian) field theory~\cite{song_gaussian_1996, martin_cavity_2008, martin_electrostatic_2008, levy_dielectric_2012} or mode coupling theory, which involves the intermediate scattering functions such as $F_{qq}(k,t)$ (see \emph{e.g.} Ref.~\cite{roy_mode_2015}). The effect of interfaces (around a solute, or at the air-water interface) on these fluctuations was also investigated by the same approaches~\cite{noah-vanhoucke_fluctuations_2009, dinpajooh_free_2015, dinpajooh_dielectric_2016, dinpajooh_free_2017, seyedi_screening_2019, matyushov_dielectric_2021}

The key role of these fluctuations on electron transfer reactions in solution has also been understood by Marcus, whose pioneering work captured the effect of the solvent polarization within continuum electrostatics~\cite{marcus1956a, marcus1965a}. The fluctuations of the so-called vertical energy gap and corresponding reorganization free energy were later sampled using \emph{ab initio} and classical molecular simulations~\cite{blumberger_redox_2006} and better described in implicit-solvent theories via \emph{e.g.} molecular Density Functional Theory~\cite{jeanmairet_molecular_2019}. The same concepts were also applied to redox reactions near metallic interfaces~\cite{reed_electrochemical_2008, takahashi_polarizable_2020, limaye_understanding_2020, kim2021interfacial} (see also Section~\ref{sec:Fluctuations:Observables:Electrode}). Electric field fluctuations also play an important role in water autodissociation~\cite{geissler_autoionization_2001} as well as vibrational dephasing in water~\cite{eaves_electric_2005} and ion pair dissociation~\cite{geissler1999a, ballard_toward_2012, kattirtzi2017a}.

The dynamics of the electric field gradient (EFG) tensor $V_{\alpha\beta}$  can be probed in NMR relaxometry experiments of quadrupolar nuclei (those with spin $I \geq 1$, such as
\textsuperscript{7}Li\textsuperscript{+},
\textsuperscript{23}Na\textsuperscript{+},
\textsuperscript{25}Mg\textsuperscript{2+},
\textsuperscript{39}K\textsuperscript{+}, 
etc.), as the coupling between the quadrupolar moment of the nucleus $eQ$ with $V_{\alpha\beta}$ usually dominates the relaxation if present~\cite{abragam1961principles}. Provided that the magnetic field $\vec{B}$ points in the $z$-direction of the laboratory frame and that the extreme narrowing regime holds (\emph{i.e.}, the typical time scale of EFG fluctuations is much smaller than the inverse Larmor frequency of the nucleus), the longitudinal relaxation rate $1/T_1$ of a quadrupolar solute can be expressed as~\cite{engstrom_molecular_1984, badu_quadrupolar_2013, carof_accurate_2014, carof_microscopic_2015, carof_collective_2016, philips_quadrupolar_2017, philips_quadrupolar_2020, mohammadi_nuclear_2020}
\begin{equation}
\label{eq:relaxation_rates}
\frac{1}{T_1} = \frac{3}{8} \frac{2I+3}{I^2(2I-1)} \left( \frac{eQ} {\hslash} \right)^2 (1+\gamma_{\infty})^2 \int_0^{+\infty} {\rm d}t \left\langle V_{zz}(t) V_{zz}(0) \right\rangle
\, .
\end{equation}
For the sake of simplicity, in Eq.~\ref{eq:relaxation_rates} we assume that the electron cloud contribution to the EFG at the nucleus can be incorporated via the Sternheimer (anti-)shielding factor~\cite{Sternheimer1950}, $\gamma_{\infty}$, and thus $V_{zz}$ is the $zz$ component of the EFG tensor obtained using the classical charge distribution around the solute. However, improvements upon the Sternheimer approximation are necessary to obtain better predictions for the quadrupolar NMR relaxation rates in aqueous electrolyte solutions~\cite{chubak_nmr_2021, chubak_quadrupolar_2023}. Quadrupolar relaxation in ionic liquids has also recently been investigated using molecular simulations~\cite{gimbal-zofka_simulations_2022}.

The relevant autocorrelation function
$\left\langle V_{zz}(t) V_{zz}(0) \right\rangle$ can be related to the charge-charge intermediate scattering function $F_{qq}(\vec{k}, t)$ of the electrolyte, see Eq.~\ref{eq:DefFzz}. As shown by Perng and Ladanyi \cite{Perng1998}, 
\begin{equation}
\label{eq:efg_dielectric}
\left\langle V_{zz}(t) V_{zz}(0) \right\rangle \approx \int_0^{+\infty} {\rm d}k \, W(k) F_{qq}(\vec{k}, t),
\end{equation}
where the weight function $W(k) = \frac{8}{5} \frac{j_1^2(ka)}{a^2}$ takes into 
account the finite solute radius $a$,
and $j_1(x)$ is the spherical Bessel function of the first kind. Note that the dielectric theory of Perng and Ladanyi~\cite{Perng1998} relies on a series of approximations: $(i)$ the solute motion is ignored; $(ii)$ a cavity construction is used to account for the finite ion size $a$; $(iii)$ translational symmetry of the electrolyte is imposed, \emph{i.e.} charge density fluctuations around the solute are assumed to be equal to those in the bulk. While such approximations oversimplify microscopic details of the solvation dynamics~\cite{carof_microscopic_2015, carof_collective_2016}, Eq.~\ref{eq:efg_dielectric} provides a straightforward way of relating the relaxation of the electric field gradient fluctuations with that of collective dielectric modes of the solution.

\subsubsection{Charge induced on an electrode}
\label{sec:Fluctuations:Observables:Electrode}

The electric fluctuations recorded with electrodes in nanoelectrochemical devices have been correlated to the microscopic dynamics of the electrolyte~\cite{zevenbergen_electrochemical_2009,mathwig_electrical_2012}. Electrode surfaces polarize in response to the presence of external charges, resulting in an induced charge density at the surface of the metal. The relation between the electrode response and the external charge density can be expressed by means of Green's functions, taking into account appropriate boundary conditions~\cite{girotto_simulations_2017}. The link between electrode response and the external charge density distribution suggests that the charge induced to the electrodes can be used to infer static and dynamical properties of the electrolyte. 

Several strategies have been used to model induced charges in molecular simulations~\cite{scalfi_molecular_2021} and to understand how they depend on electrolyte configurations~\cite{geada_insight_2018, pireddu_molecular_2021}. Molecular simulations of nanocapacitors (viz. electrolyte confined between two polarizable electrodes) have been used to estimate electrochemical properties from the fluctuations of induced charges, and to understand how they relate to the behaviour of the confined electrolyte. The differential capacitance $C_{\rm diff}$ of a capacitor, \emph{i.e.} the derivative of the average charge on the electrodes with respect to the applied voltage $\Delta\Psi$ between them, can be estimated from the fluctuations of the total electrode charge~\cite{johnson1928a, nyquist1928a, limmer_charge_2013, scalfi_charge_2020}
\begin{equation} \label{eq:capcharge}
    C_{\rm diff}= \frac{\partial \langle Q \rangle}{\partial \Delta\Psi} = \beta \langle \delta Q^2 \rangle \, ,
\end{equation}
where $\delta Q = Q - \langle Q \rangle$. Furthermore, the frequency-dependent admittance $Y(\omega)$ can be estimated from the dynamical fluctuations of the electrode charge, using the following fluctuation-dissipation relation~\cite{pireddu_frequency_2022}
\begin{equation} \label{eq:admittancecharge}
    Y(\omega) =  \beta \left[ i\omega \langle \delta Q^2 \rangle + \omega^2 \int_{0}^{\infty}\langle \delta Q(0) \delta Q(t) \rangle e^{-i\omega t}{\rm d}t \right] .
\end{equation}
In the absence of applied voltage, the total charge induced on the electrodes is proportional to the total polarization of the electrolyte in the direction perpendicular to the electrode surfaces. At finite voltage, the relation between electrode charge and polarization also involves a contribution proportional to the magnitude of the external field~\cite{takae_fluctuations_2015}. This allows rewriting Eqs.~\ref{eq:capcharge} and~\ref{eq:admittancecharge} using the electrolyte polarization, thus creating a direct connection between the electrochemical properties of nanocapacitors and the microscopic fluctuations of the electrolytes.

\section{The charge-charge dynamic structure factor of aqueous electrolyte solutions}
\label{sec:Szz}

We now illustrate the above discussion for the specific case of an aqueous sodium chloride solution. We investigate the charge-charge dynamic structure factor for an electrolyte described with explicit ions and solvent molecules using molecular dynamics (MD) simulations, as well as with explicit ions in an implicit solvent characterized by a dielectric constant $\epsilon_r$ using Langevin dynamics (LD) and Brownian dynamics (BD) simulations. We also consider the predictions of theories corresponding to the implicit solvent description. The theory and simulation details are presented in Sections~\ref{sec:Szz:Theory} and~\ref{sec:Szz:Simulation}, respectively. All results are reported and discussed in Section~\ref{sec:Results}.

\subsection{Theory}
\label{sec:Szz:Theory}

\subsubsection{Poisson-Nernst-Planck}
\label{sec:Szz:Theory:PNP}

A standard theory for the dynamics of electrolyte solutions is the Poisson-Nernst-Planck (PNP) model~\cite{hunter_book}. Despite its simplicity and its limitations as the salt concentration increases, it captures the basic ingredients of the ionic dynamics, namely thermal diffusion and the effect of electrostatic interactions, which are treated at the mean-field level. It combines a conservation (Nernst-Planck) equation for the local density $\rho_{\alpha}(\vec{r},t)$ of ionic species with charge $q_\alpha=z_\alpha e$ and diffusion coefficient $D_\alpha$:
\begin{align}
    \frac{\partial \rho_\alpha}{\partial t} + \nabla\cdot\left[ -D_\alpha \nabla \rho_\alpha - \beta D_\alpha q_\alpha\rho_\alpha \nabla \phi \right] &= 0 
    \, ,
    \label{eq:NernstPlanck}
\end{align}
with the Poisson equation satisfied by the electrostatic potential in the implicit solvent:
\begin{align}
    \Delta \phi = -\frac{\rho_q}{ \epsilon_0\epsilon_r }
    \, .
    \label{eq:Poisson}
\end{align}
Analytical results can be obtained by considering small deviations of the concentrations and potential around the average values, $\rho_\alpha(\vec{r},t)=\rho_\alpha^0+\delta\rho_\alpha(\vec{r},t)$, and linearizing the PNP equations. At equilibrium, one recovers the Debye-H\"uckel (linearized Poisson-Boltzmann) solution, where the potential and density perturbations decay over the characteristic Debye screening length $\lambda_D=\kappa_D^{-1}$, with
\begin{align}
    \kappa^2_D = 4 \pi l_B \sum_\alpha \rho_\alpha^0 z_\alpha^2
    \label{eq:KappaD}
\end{align}
and the Bjerrum length $l_B=\beta e^2/4\pi\epsilon_0\epsilon_r$. Relaxation of charge fluctuations occurs over the so-called Debye relaxation time 
\begin{align}
   \frac{1}{\tau_D} = 4 \pi l_B \sum_\alpha \rho_\alpha z_\alpha^2 D_\alpha
   \, .
   \label{eq:TauD}
\end{align}

While analytical results can be obtained in the general case where cations and anions have different diffusion coefficients, in the following we will discuss only the simpler one where both are equal, $D_+=D_-=D$, and use the average value between that of Na$^+$ and Cl$^-$ ions for the comparison with simulations. The difference in diffusion coefficients induces an internal electric field, a process resulting at long times in the common diffusion of the ions (Nernst-Hartley)~\cite{robinson_book}. The result for the average diffusion coefficient $D=(D_+ + D_-)/2$ neglects corrections of order  $|D_+ -D_-|/|D_++D_-|$. Under this simplifying assumption, the Debye time Eq.~\ref{eq:TauD} reduces to $\tau_D  = 1/D \kappa_D^2$, and the quantities defined in Section~\ref{sec:Fluctuations:DensityCurrent} can be computed analytically:
\begin{align}
    S_{qq}^\textrm{PNP}(k) &= \frac{k^2}{k^2+\kappa_D^2} 
    \, ,
    \label{eq:SzzStaticPNP} \\
    F_{qq}^\textrm{PNP}(k,t) &= S_{qq}^\textrm{PNP}(k) e^{-D(k^2+\kappa_D^2)t }  
    \, ,
    \label{eq:FzzPNP} \\
    \tilde{F}_{qq}^\textrm{PNP}(k,\omega) &= \frac{S_{qq}^\textrm{PNP}(k)}{ -i \omega + D(k^2+\kappa_D^2) }  
    \, ,
    \label{eq:tildeFzzPNP} \\
    S^\textrm{PNP}_{qq}(k,\omega) &= \frac{2 \,Dk^2}{\omega^2+D^2(k^2+\kappa^2_D)^2}
        \, .
    \label{eq:SzzPNP}
\end{align}

\subsubsection{Dynamical Density Functional Theory}
\label{sec:Szz:Theory:DDFT}

The above results can be slightly generalized to go beyond some limitations of the PNP model, which treats the electrolyte as a gas of ions interacting only via the mean-field electrostatic potential. This model is as a special case of Dynamical Density Functional Theory (DDFT). The following steps can be found in more detail in the recent review by te Vrugt \emph{et al.}~\cite{te_vrugt_classical_2020}. DDFT models the evolution of a density field $\rho(\vec{r},t)$ as:
\begin{align}
    \frac{\partial \rho}{\partial t} &= \nabla  \cdot \left( \beta D \cdot \rho \, \nabla \frac{\delta \mathcal{F[\rho]}}{\delta \rho}\right)
    \, ,
    \label{eq:DDFT}
\end{align}
where the right-hand side comes from the divergence of a flux involving a density-independent mobility $\beta D$ and the gradient of a local chemical potential, which is the functional derivative of the free energy functional $\mathcal{F}[\rho]$. The latter can be decomposed into an ideal term leading to the usual ideal part of the chemical potential, and an excess term arising from the interactions. The PNP model is recovered by describing the excess term as the mean-field electrostatic energy, but it is possible to include more elaborate models to capture \emph{e.g.} some of the steric and electrostatic correlations between the ions. As in the PNP case, one can then linearize the density around a homogeneous state with density $\rho_0$ to arrive at the following evolution equation for the excess density $\delta\rho=\rho-\rho_0$:
\begin{align}
    \frac{\partial \delta\rho}{\partial t} =  D \Delta \delta \rho - D \rho_0 \Delta  \int c^{(2)}(\left|\vec{r}-\vec{r}'\right|;\rho_0) \,\delta\rho(\vec{r}',t) \mathrm{d}\vec{r}'
    \, , 
    \label{eq:DDFTlin}
\end{align}
where $c^{(2)}(r;\rho_0)$ is the direct pair correlation function for a homogeneous fluid of density $\rho_0$, which is, up to a factor $k_BT$, the second order functional derivative of the free energy. The solution of this equation is easily found in reciprocal space, using the relation $1-\rho_0 \,\hat{c}^{(2)}(k ; \rho_0) = S(k)$ with the static structure factor (which comes from the Ornstein-Zernike equation, see Ref.~\citenum{hansen_mcdonald_theory_of_simple_liquids_book}), with the result:
\begin{align}
    \delta \hat{\rho}(k,t) = \delta \hat{\rho}(k,0) e^{-Dk^2 t /S(k)} 
    \, .
    \label{eq:DDFTsolution}
\end{align}
For the charge-charge correlation function, this leads to the extension of Eqs.~\ref{eq:FzzPNP} and~\ref{eq:SzzPNP}:
\begin{align}
    F_{qq}(k,t) \simeq S_{qq}(k) \, e^{-Dk^2 t /S_{qq}(k) }
    \label{eq:FzzDDFTapprox}    
\end{align}
and
\begin{align}
    S_{qq}(k,\omega) \simeq \frac{2 \, D k^2}{\omega^2 + [D k^2/S_{qq}(k)]^2}
    \, .
    \label{eq:SzzDDFTapprox}    
\end{align}
This result can be used to model the dynamics of the charge distribution beyond PNP if a more accurate model of the structure is available, either from improved free energy functionals, or from simulations. We will consider the latter strategy in Section~\ref{sec:Results:ExplicitImplicit}. One should however keep in mind that these results also rely on the linearization of the density around its average and neglect features of the dynamics that may play a role, such as hydrodynamic effects. We also note that other forms of DFT, in particular stochastic DFT (with various approximations), have recently been used to predict the conductivity of electrolyte, including thermal noise and in some cases hydrodynamic effects, both at equilibrium and for nonequilibrium steady-states in the presence of an applied electric field, in the bulk and under confinement~\cite{zorkot_current_2016, zorkot_power_2016, zorkot_current_sPNP_2018, mahdisoltani_long_2021, mahdisoltani_transient_2021, demery_conductivity_2016, peraud_fluctuating_2017, donev_fluctuating_hydro_2019, avni_conductivity_2022, avni_conductance_2022, bonneau_temporal_arxiv_2023}.

\subsubsection{Ballistic regime}
\label{sec:Szz:Theory:Ballistic}

In the case of an explicit solvent, following Newton's laws of motion, the behaviour at high frequency ($\omega\to\infty$) and small distances ($k\to\infty)$ can be modelled as a ballistic regime, provided that the probed distances are sufficiently short (typically a fraction of the distance between ions and molecules, which is also comparable to their size). In that case, one can also neglect correlations between ions, so that the correlation of the sum over ions reduces to a sum of self terms. The resulting structure factor can be understood as the Maxwell-Boltzmann distribution of velocities $\omega/k$ for each ion~\cite{hansen_mcdonald_theory_of_simple_liquids_book}:
\begin{align}
    S^\textrm{Ball.}_{qq}(k,\omega) &= \frac{1}{2}\left[ 
    \sqrt{\frac{2\pi\beta m_+}{k^2}} e^{- \beta m_+ \omega^2 / 2 k^2} + 
    \sqrt{\frac{2\pi\beta m_-}{ k^2}} e^{- \beta m_- \omega^2 /2 k^2} \right]
    \, ,
    \label{eq:SzzBall}
\end{align}
where $m_+$ and $m_-$ are the masses of the cations and anions, respectively.

\subsection{Simulation details}
\label{sec:Szz:Simulation}

We simulate a bulk aqueous sodium chloride solution using MD, LD and BD simulations, all using periodic boundary conditions. For MD, the simulated system consists of $N_{\text{water}}=3050$ water molecules and $N_{\text{NaCl}}=70$ ion pairs. Water molecules are described using the SPC/E water model~\cite{berendsen1987missing} and ions by the Joung-Cheatham force-field~\cite{mester2015mean}. Short-range Lennard-Jones interactions between unlike particles are computed using the Lorentz-Berthelot mixing rules; they are truncated and shifted at a cut-off $r=9$~\AA. Long-range electrostatic interactions are computed with the PPPM method~\cite{hockney1988computer}. Newton's equations of motion are integrated with the velocity Verlet algorithm using a timestep $\delta t=2$~fs, and water molecules are treated as rigid using the SHAKE algorithm~\cite{andersen1983rattle}. The ionic solutions is first equilibrated at $p=1$~atm and $T=298.15$~K using the Nos\'{e}-Hoover barostat and thermostat, with time constants of 1000~fs and 100~fs, respectively. The resulting average box size $L_{box}=45.6$ \AA, corresponding to a salt concentration of 1.23~M, is then used for $NVT$ simulations at the same temperature. Three independent simulations are run, for at least 20~ns of equilibration followed by 80~ns of production. Each production run is divided into 6 blocks considered as independent for the analysis. We also perform simulations for pure water using the same procedure and the same number of water molecules, as well as simulations at infinite dilution with a single cation and anion to determine their diffusion coefficients from the slope of their mean-square displacement. The static permittivity $\epsilon_s$ is calculated from the fluctuations of the total water dipole moment $\vec{M}_W$ in the aqueous solutions and pure water as~\cite{caillol_dielectric_1987}
\begin{align}
    \epsilon_s &= 1+\frac{\beta}{3\epsilon_0V}\langle\delta \vec{M}_W^2\rangle 
    \,,
    \label{eq:CalcEpsilon}
\end{align}
where $V$ is the volume of the simulation box.

For implicit solvent simulations (LD and BD), short-range interactions between ions are identical to that for MD simulations, described above. The Coulomb interactions are still computed using the PPPM method, but screened by the dielectric constant of the solvent. We use the experimental value~\cite{Malmberg1956Dielectric} $\epsilon_r = 78.5$, which is slightly larger than the one of the SPC/E water model ($\epsilon_r = 70.5$)~\cite{sanchez2019uncertainties}. The diffusion coefficients $D_\pm$ of the ions for BD, or corresponding friction coefficients for Lanvegin dynamics ($\gamma_\pm=1/\beta D_\pm$), are taken from MD simulations at infinite dilution as described above, namely: $D_{\textrm{Na}^+} = 1.54~10^{-9}$~m$^2$.s$^{-1}$ and $D_{\textrm{Cl}^-} = 1.28~10^{-9}$~m$^2$.s$^{-1}$. The equations of motion are integrated using the velocity Verlet algorithm with a timestep $\delta t = 2$~fs coupled with a Langevin thermostat fixed at $T = 298.15$~K for the underdamped LD, and the overdamped BAOAB integrator~\cite{Leimkuhler_BAOAB_2013} with $\delta t = 25$~fs for BD. These simulations are much less computationally demanding than MD, so that we study the same salt concentration of $1.23$~M using a larger system with  $N_{\rm{NaCl}}=560$ ion pairs in a cubic box of size $2 L_{\rm{box}}$, for much longer simulations times. For both LD and BD, we perform five independent runs. The total simulation time for each production run is 5~$\mu$s for LD and 50~$\mu$s for BD. Each of them is then divided into 50 blocks considered as independent for the analysis.

All MD, LD and BD simulations are performed with the LAMMPS simulation package~\cite{thompson_LAMMPS_2022}. The Fourier components of the charge density, $\hat{\rho}_{q}(\vec{k},t)$ (Eq.~\ref{eq:RhoZofk}), are sampled every $6$~fs, $400$~fs, and $1$~ps for MD, LD and BD respectively, for selected wavevectors compatible with the periodic boundary conditions, satisfying $|\vec{k}|=n k_{min}$ with $k_{min}=2\pi/L_{box}$, where $L_{box}$ is the box size for MD simulations and $n$ integers between 1 and 256.

This covers length scales ranging between 0.2 and 45.6~\AA, smaller than the particle size and larger than the typical correlation lengths in the electrolyte. For LD and BD simulations, with a box size $2L_{box}$, we also consider $|\vec{k}|= k_{min}/2$. The correlation function $F_{qq}(\vec{k},t)$ and its Fourier transform $S_{qq}(\vec{k},\omega)$ are then computed from the time series of $\hat{\rho}_{q}(\vec{k},t)$ using fast Fourier transforms (FFT). In the absence of external field the three directions of space are equivalent: In order to improve the statistics we consider wavevectors in the $x$, $y$ and $z$ directions and average the results. The reported results further correspond to averages over the runs and blocks, with uncertainties estimated as the standard error between independent realizations.

\section{Results}
\label{sec:Results}

Section~\ref{sec:Results:ExplicitImplicit} first examines the ionic contribution to the charge fluctuations, comparing the results from MD simulations in the presence of an explicit solvent, with LD and BD using an implicit solvent, as well as theoretical predictions described in Section~\ref{sec:Szz}.
Section~\ref{sec:Results:IonsWater} is then devoted to the contributions of ions and water to the total charge fluctuations.

\subsection{Ionic contribution to the charge fluctuations}
\label{sec:Results:ExplicitImplicit}

\begin{figure}[ht!]
  \includegraphics[width=0.45\textwidth]{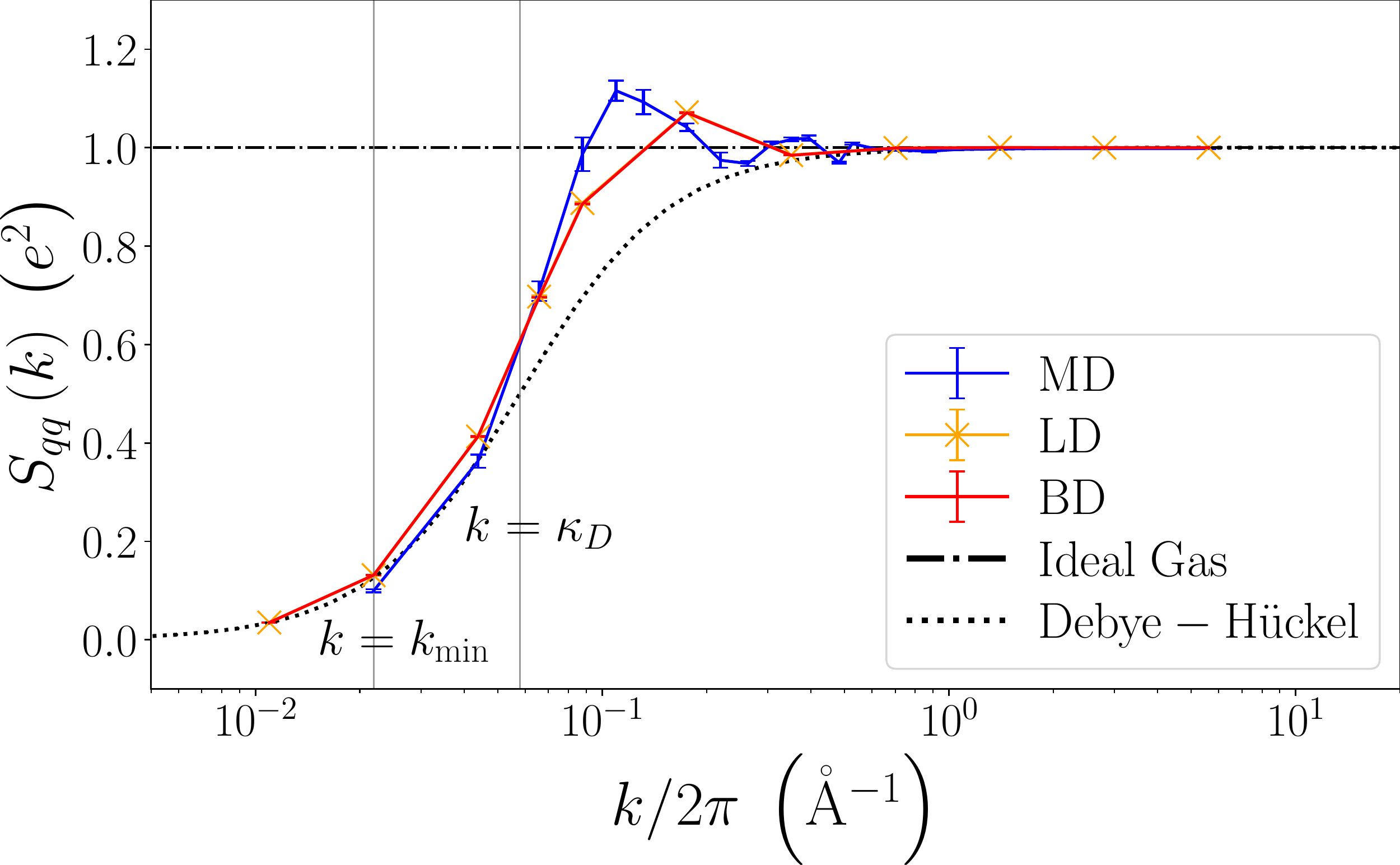}
    \caption{
    Static charge-charge structure factor $S_{qq}(k)$ (see Eq.~\ref{eq:DefSzzStatic}), including only ions. The figure displays results from molecular dynamics (MD, blue), Langevin dynamics (LD, yellow), and Brownian dynamics (BD, red) simulations, for wave vectors ranging from the minimal value for the box size of the MD simulations, $k_{min}=2\pi / L_{box}$, to $256 k_{min}$, corresponding to wavelengths between $45.6$ and $0.2$~\AA, as well as for $k_{min}/2$ in the LD and BD case for which a larger simulation box was used. The results are also compared with the prediction of PNP theory (see Eq.~\ref{eq:FzzPNP}, dashed line) and for an ideal gas (dashed-dotted line). The vertical dotted lines indicate $k=k_{min}$ and $k=\kappa_D$, the inverse Debye screening length.
    }
    
    \label{fig:StaticSzzComparison}
    
\end{figure}

Fig.~\ref{fig:StaticSzzComparison} shows the static charge-charge structure factor $S_{qq}(k)=F_{qq}(\vec{k},t=0)$ for the various levels of description. For sufficiently large $k>10$~\AA$^{-1}$, corresponding to wavelengths shorter than the ionic size (hence distances much shorter than the typical distance between ions), all results converge to the ideal gas result $S_{qq}(k)=1$. For sufficiently small $k\lesssim\kappa_D$, the MD, LD and BD results are similar and well described by the linearized mean-field Debye-H\"uckel theory. Despite the limited accessible range of wavevectors, the simulations seem to follow the corresponding scaling as $k^2$ (see Eq.~\ref{eq:SzzStaticPNP}), which reflects the screening of electric fields by the ions (see also the discussion of the Stillinger-Lovett conditions in Section~\ref{sec:Results:IonsWater}). Nevertheless, Debye-H\"uckel theory is not expected to be quantitative for relatively high salt concentrations (typically, beyond $10^{-2}$~M) such as the one considered here. This is evident for intermediate $k$, where even the results of implicit solvent simulations are not recovered, suggesting the importance of ion-ion correlations, which may be both of electrostatic and steric origin, in this range~\cite{leote_de_carvalho_decay_1994, coupette_screening_2018}. In addition, the BD results also deviate from the MD ones. This highlights the (expected) shortcomings of the underlying implicit solvent model for lengths scales comparable to the size of ions and water molecules.

\begin{figure}[ht!]
\centering
    \includegraphics[width=0.43\textwidth]{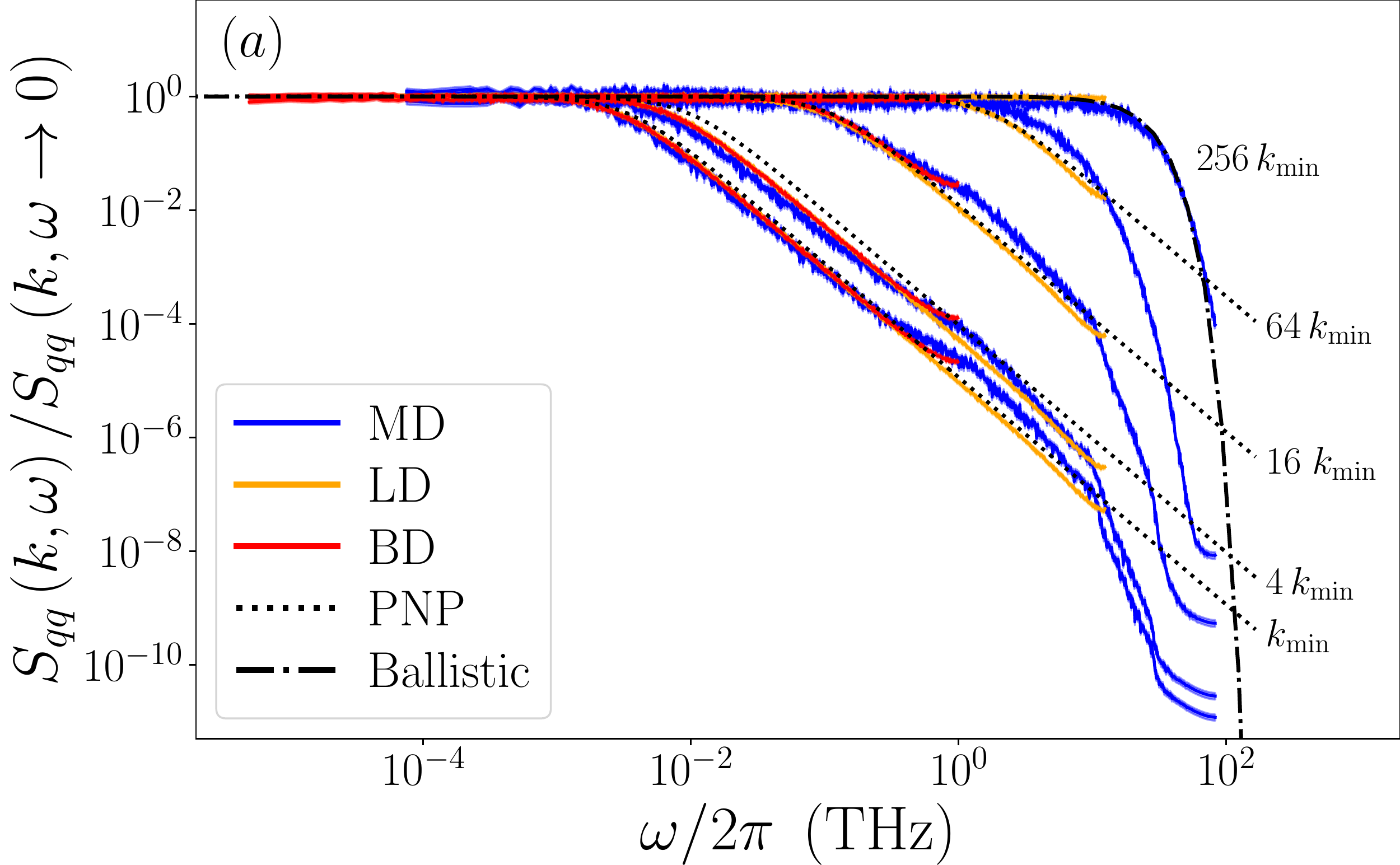}\\ \vspace{0.2cm}
    \includegraphics[width=0.43\textwidth]{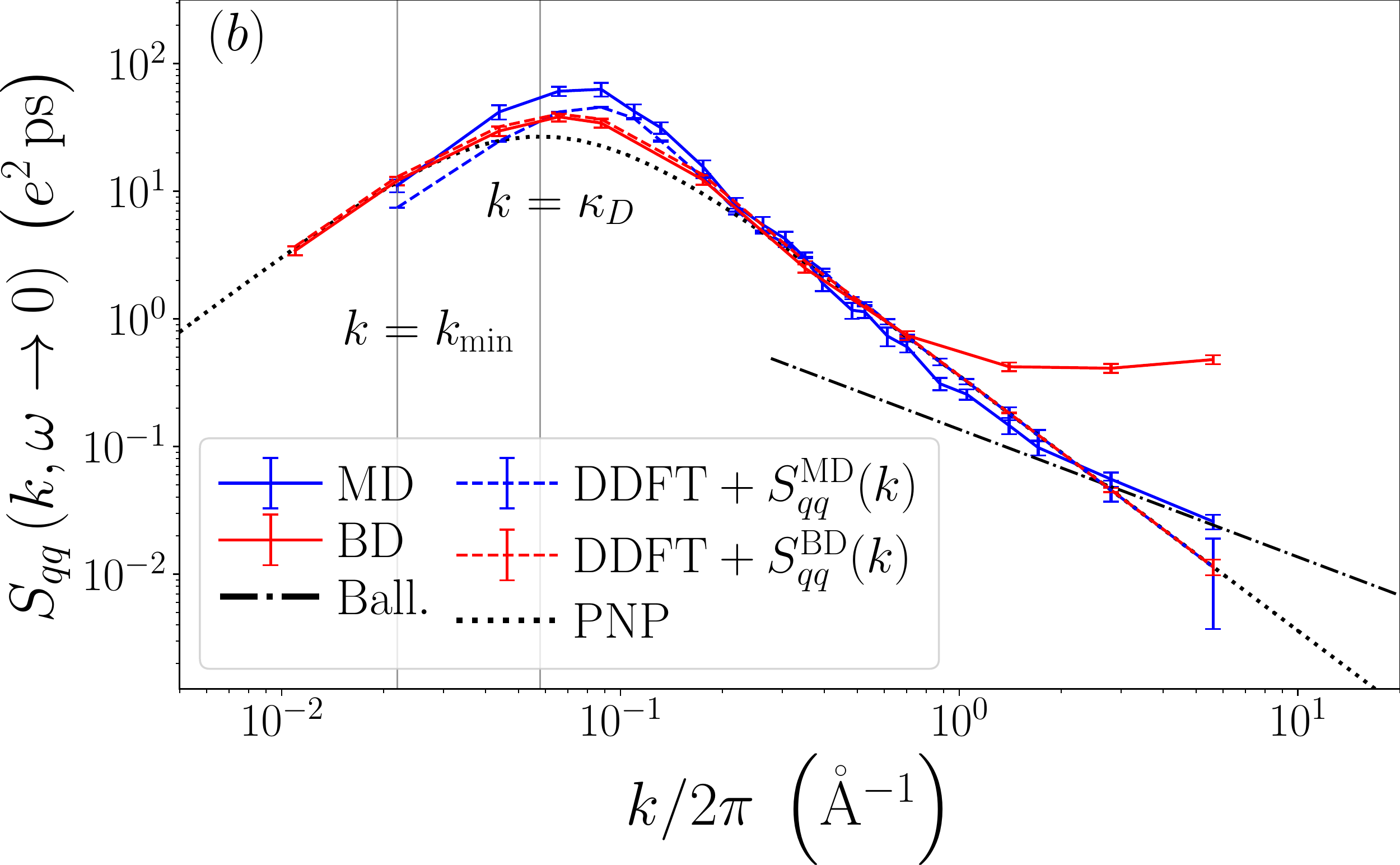}\\ \vspace{0.2cm}
    \includegraphics[width=0.43\textwidth]{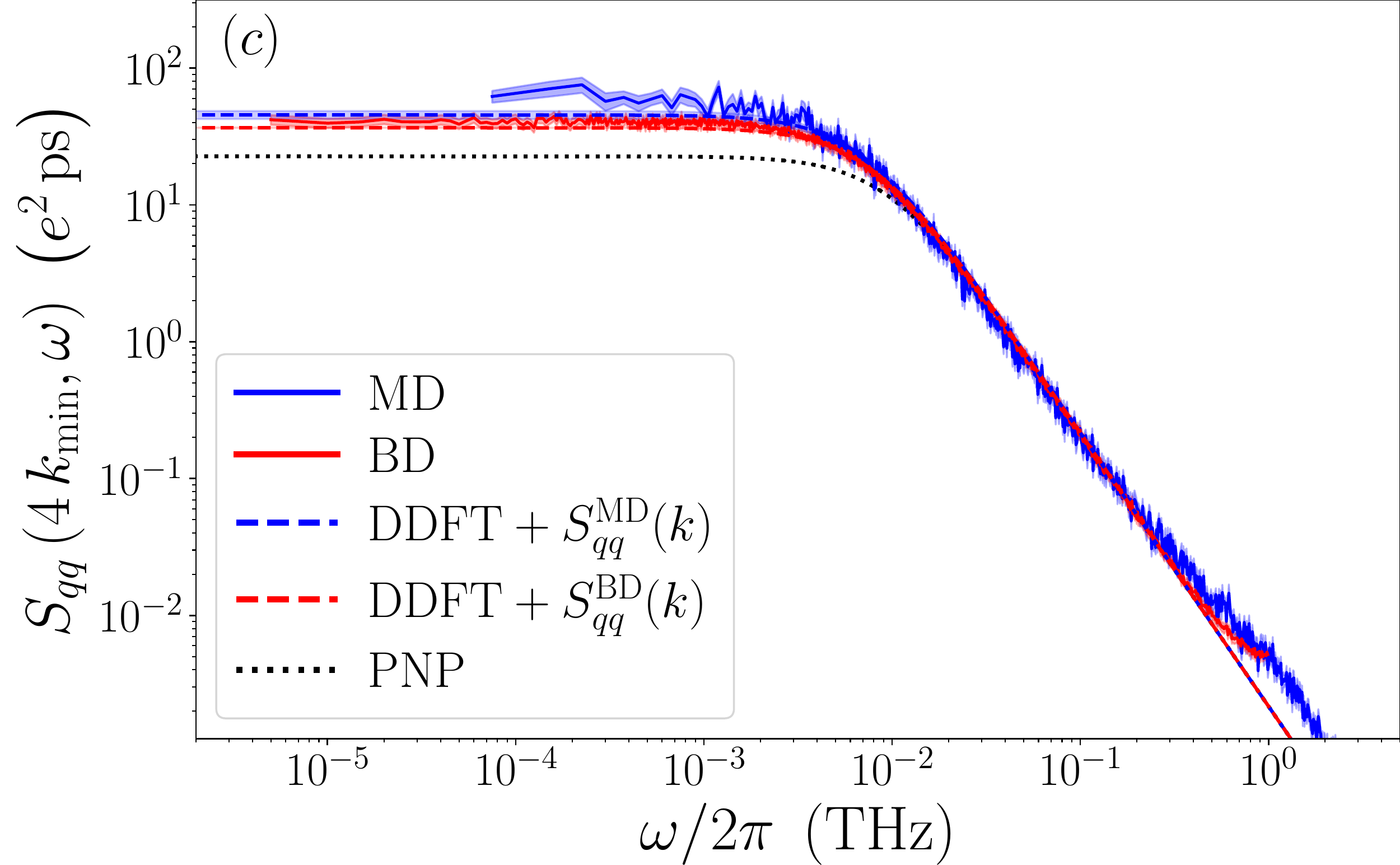}
    \caption{
    (a) Dynamic charge-charge structure factor $S_{qq}(\vec{k},\omega)$ (see Eq.~\ref{eq:DefSzzofkandomega}), including ions only, normalized by its initial value $S_{qq}(\vec{k},\omega=0)$, which is reported in panel (b). 
    Both panels show results from molecular dynamics (MD, blue), Langevin dynamics (LD, yellow, in panel (a) only), and Brownian dynamics (BD, red) simulations, for wave vectors ranging from the minimal value for the box size of the MD simulations, $k_{min}=2\pi / L_{box}$, to $256 k_{min}$, corresponding to wavelengths between $45.6$ and $0.2$~\AA,  as well as for $k_{min}/2$ in the LD and BD case for which a larger simulation box was used. (c) Dynamic charge-charge structure factor for $k=4k_{min}$, from MD and BD. The simulation results are also compared with the prediction of Dynamical Density Functional Theory Eq.~\ref{eq:SzzDDFTapprox} (DDFT, dashed lines), using the static structure factor $S_{qq}(k)$ obtained in the corresponding simulations in panels (b) and (c), or from Debye-H\"uckel theory (\emph{i.e.} PNP, see Eq.~\ref{eq:SzzPNP}, black dotted line) in all panels. In panels (b) and (c), the MD predictions are scaled by the appropriate ratio of number of atoms for comparison with the other models (see text). Panels (a) and (b) also show the prediction for the ballistic regime (Eq.~\ref{eq:SzzBall}, dashed-dotted line). The vertical dotted lines in panel (b) indicate $k=k_{min}$ and $k=\kappa_D$, the inverse Debye screening length.
    }
    \label{fig:SzzComparison}
\end{figure}

Fig.~\ref{fig:SzzComparison}a then displays the charge-charge dynamic structure factor $S_{qq}(k,\omega)$ defined by Eq.~\ref{eq:DefSzzofkandomega}, including ions only, as a function of frequency $\omega$ and normalized by its initial value. The results are shown from MD, LD and BD simulations for selected wave vectors ranging from the minimal value for the box size of the MD simulations, $k_{min}=2\pi / L_{box}$, to $256 k_{min}$, corresponding to wavelengths between $45.6$ and $0.2$~\AA. The MD predictions are scaled by the appropriate ratio of number of atoms, $2N_\text{NaCl}/(3N_\text{water}+2N_\text{NaCl})$, since these numbers enter in the definition of the charge-charge dynamic structure factor (see Eq.~\ref{eq:DefFzz}). We also report the predictions of PNP theory (Eq.~\ref{eq:SzzPNP}) and, for the largest wavenumbers, for the ballistic regime (Eq.~\ref{eq:SzzBall}). 

For all wave vectors and levels of description, $S_{qq}(\vec{k},\omega)$ decays from its initial value to 0 for $\omega\to\infty$, but the crossover occurs at higher frequencies for increasing $k$, reflecting a faster decorrelation of the charge over shorter length scales. Furthermore, the shape of the decay depends on $k$ and on the level of description. For the largest $k=256 k_{min}$, the MD results follow the Gaussian decay predicted by Eq.~\ref{eq:SzzBall}, without any adjustable parameter. This ballistic regime, expected when the wavelength is shorter than the distance between ions and molecules, is not captured by the implicit solvent models. As $k$ decreases, $S_{qq}(\vec{k},\omega)$ gradually changes toward a limiting curve (corresponding to the $k\to0$ limit, even though it cannot be reached for a finite box size), which displays a Lorentzian shape over a wide frequency range. Some features of the MD results, especially at high frequency, are not reproduced by the implicit solvent simulations, which neglect the details of the short-time dynamics, modeled only by the friction force. Nevertheless, LD and BD simulations correctly capture the behaviour at lower frequency, both in terms of shape and crossover frequency, which grows with increasing $k$. This confirms the relevance of the choice of friction for LD and BD, which was determined from the diffusion coefficients in MD simulations, as explained in Section~\ref{sec:Szz:Simulation}. Of course, since the time step for BD is larger than in MD, the largest frequency that can be sampled with such simulations does not reach that of the latter, but they can be used to probe lower frequencies.

The Lorentzian decay predicted by PNP theory (Eq.~\ref{eq:SzzPNP}) corresponds to the diffusion of charge over a distance $1/\sqrt{k^2+\kappa_D^2}$, which grows with decreasing $k$ and reduces in the limit $k\ll\kappa_D$ to the Debye length $\lambda_D\sim 0.364$~\AA. The corresponding crossover frequency of $D\kappa_D^2$, the inverse of the Debye time. The predictions are in semi-quantitative agreement with the LD and BD simulations, but the observed crossover frequency is slightly higher than the one from simulations. Several factors can contribute to such a discrepancy, in particular the fact that the static correlations between ions are not well described at such a high salt concentration, as discussed above. In addition, one should keep in mind that for the high concentration considered here, the Debye length is shorter than the ionic size, so that this is not the most relevant correlation length~\cite{leote_de_carvalho_decay_1994, janecek_effective_2009, lee_scaling_2017, lee_underscreening_2017, rotenberg_underscreening_2018, coupette_screening_2018, coles_correlation_2020, krucker-velasquez_underscreening_2021, zeman_ionic_2021, cats_primitive_2021}. From the more general DDFT approach, Eq.~\ref{eq:SzzDDFTapprox}, one predicts that the crossover frequency is given in the $k\to0$ limit by $\lim_{k\to0} Dk^2/S_{qq}(k)$, which reduces to $D\kappa_D^2$ when the free energy functional corresponds to Debye-H\"uckel theory.

The initial value of the dynamic charge-charge structure factor $S_{qq}(\vec{k},\omega=0)$ shown in Fig.~\ref{fig:SzzComparison}b also illustrates many of the similarities and differences between levels of description discussed so far on the frequency-dependence. The BD simulations reproduce correctly the MD results at small and large $k$ but fail to capture the intermediate range, where the maximum is rather well located (close to $\kappa_D$) but underestimated. PNP theory also captures these two limiting regimes and only qualitatively captures the BD results in the intermediate range, with a maximum predicted at $k=\kappa_D$ but underestimated compared to BD. Introducing the static structure factor $S_{qq}(k)$ from BD in the DDFT result Eq.~\ref{eq:SzzDDFTapprox} significantly improves the predictions. This indicates that most of the limitations of PNP theory follow from that of Debye-H\"uckel theory to predict the structure at the relatively large concentration considered here (see Fig.~\ref{fig:StaticSzzComparison}). In contrast, for MD it is not sufficient to introduce $S_{qq}(k)$ in Eq.~\ref{eq:SzzDDFTapprox} to recover $S_{qq}(\vec{k},\omega=0)$. 

The same observations can be made for non-zero frequencies, as illustrated for $k=4k_{min}=0.55$~\AA$^{-1}$\ in Fig.~\ref{fig:SzzComparison}c. BD correctly captures the high-frequency results of MD in the considered range (below 1~THz, significantly lower than the upper range covered in panel~\ref{fig:SzzComparison}a), and the crossover frequency toward the $\omega\to0$ value. The latter is underestimated by BD. PNP overestimates the crossover frequency and underestimates the $\omega\to0$ limit. Introducing the static structure factor from BD in Eq.~\ref{eq:SzzDDFTapprox} provides a good description of $S_{qq}(\vec{k},\omega)$, but the same procedure for MD only partly improves the results, suggesting that other effects arising from the dynamic correlations related to the explicit solvent are at play. Even though this is beyond the scope of this work, it might be possible to capture part of these effects in more advanced DDFT relaxing in particular the assumption of a density-independent mobility~\cite{te_vrugt_classical_2020}, in BD simulations, \emph{e.g.} by introducing hydrodynamic interactions~\cite{jardat_transport_1999,jardat_brownian_2000, jardat_brownian_2004}, or using other advanced mesoscopic simulation techniques taking the coupling with hydrodynamic flows into account~\cite{dahirel_hydrodynamic_2018, rotenberg_coarse-grained_2010, pagonabarraga_recent_2010, tischler_thermalized_2022}, as well in analytical theories for transport in electrolytes~\cite{dufreche_analytical_2005, contreras_aburto_unifying_2013, demery_conductivity_2016, peraud_fluctuating_2017, banerjee_ions_2019, donev_fluctuating_hydro_2019, avni_conductivity_2022}.

\subsection{Ion and water contributions to the charge fluctuations}
\label{sec:Results:IonsWater}

We now focus on the case of MD simulations with an explicit solvent to analyze the contributions of ions and water to the charge-charge dynamic structure factor. To that end, we split the charge density in Eq.~\ref{eq:RhoZofk} into two sums over Na$^+$ and Cl$^-$ ions (I) and water oxygen and hydrogen atoms (W), respectively:
\begin{align}
    \tilde{\rho}_q(\vec{k},t) &= \tilde{\rho}_q^{\text{I}}(\vec{k},t) + \tilde{\rho}_q^{\text{W}}(\vec{k},t)
    \, .
    \label{eq:rhoIandW}
\end{align}
The charge-charge dynamic structure factor defined by Eqs.~\ref{eq:DefFzz} and~\ref{eq:DefSzzofkandomega} can then be expressed as
\begin{align}
    S_{qq}^{tot}(k,\omega) &= S^{I}_{qq}(k,\omega) + S^{W}_{qq}(k,\omega) + S^{IW}_{qq}(k,\omega) 
    \, ,
    \label{eq:SzzIandW}
\end{align}
corresponding to ion-ion, water-water and cross terms. Note that the denominator in Eq.~\ref{eq:DefFzz} is $N$, the total number of atoms, for all three contributions to $\langle\hat{\rho}_q(\vec{k},t)\hat{\rho}_q(-\vec{k},0)\rangle$.

\begin{figure}[ht!]
\centering
    \includegraphics[width=0.45\textwidth]{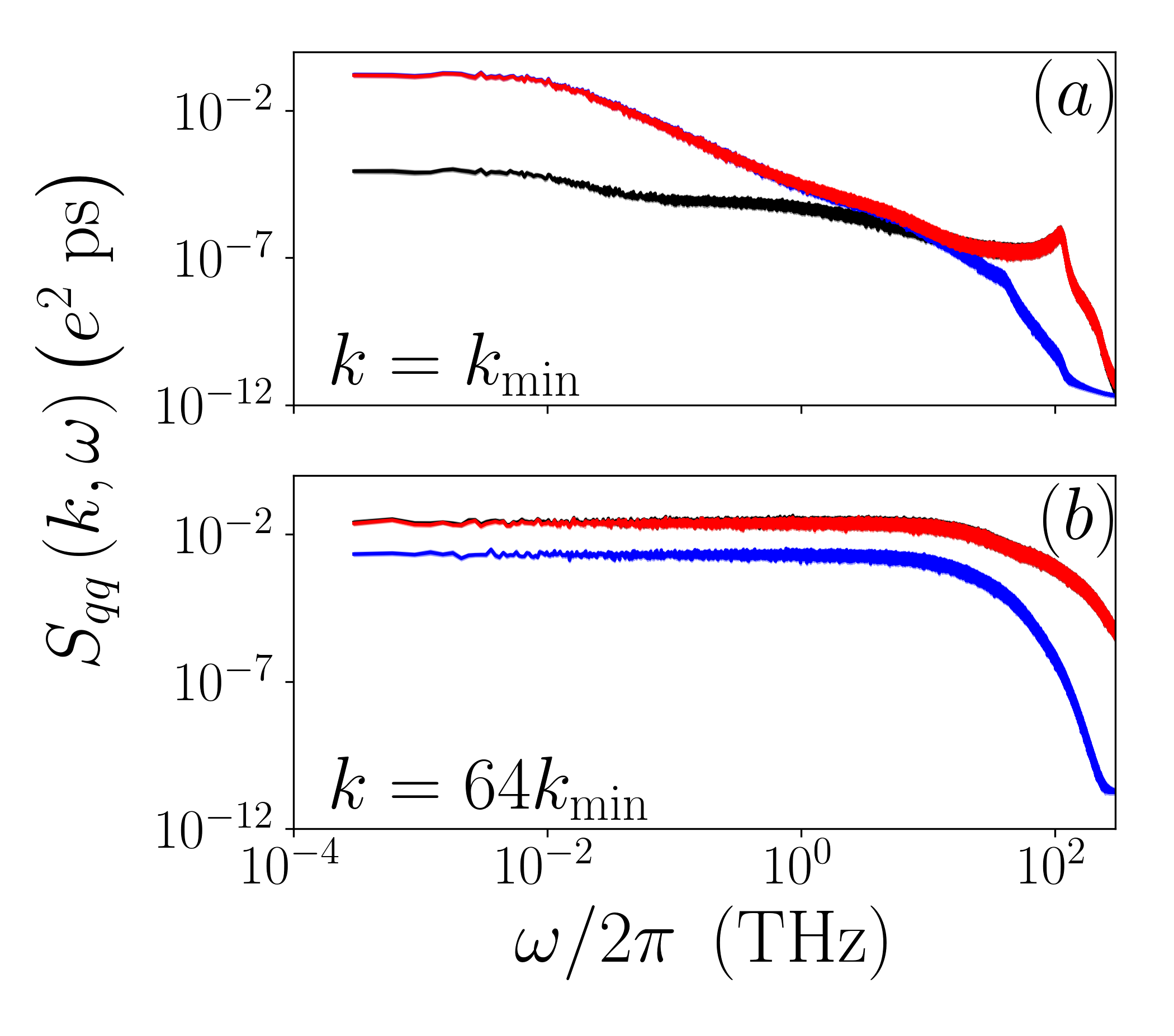}
    \caption{
    Dynamic charge-charge structure factor for the MD simulations with an explicit solvent as a function of frequency for $k=k_{min}$ (a) and $k=64k_{min}$ (b).
    Each panel shows the total $S_{qq}^{tot}(k,\omega)$ in black, as well as the contributions of ion-ion correlations, $S^{I}_{qq}(k,\omega)$, in blue and of water-water correlations, $S^{W}_{qq}(k,\omega)$, in red (see Eq.~\ref{eq:SzzIandW}). 
    }
    \label{fig:SzzContributions}
\end{figure}

\begin{figure}[ht!]
\centering
    \includegraphics[width=0.45\textwidth]{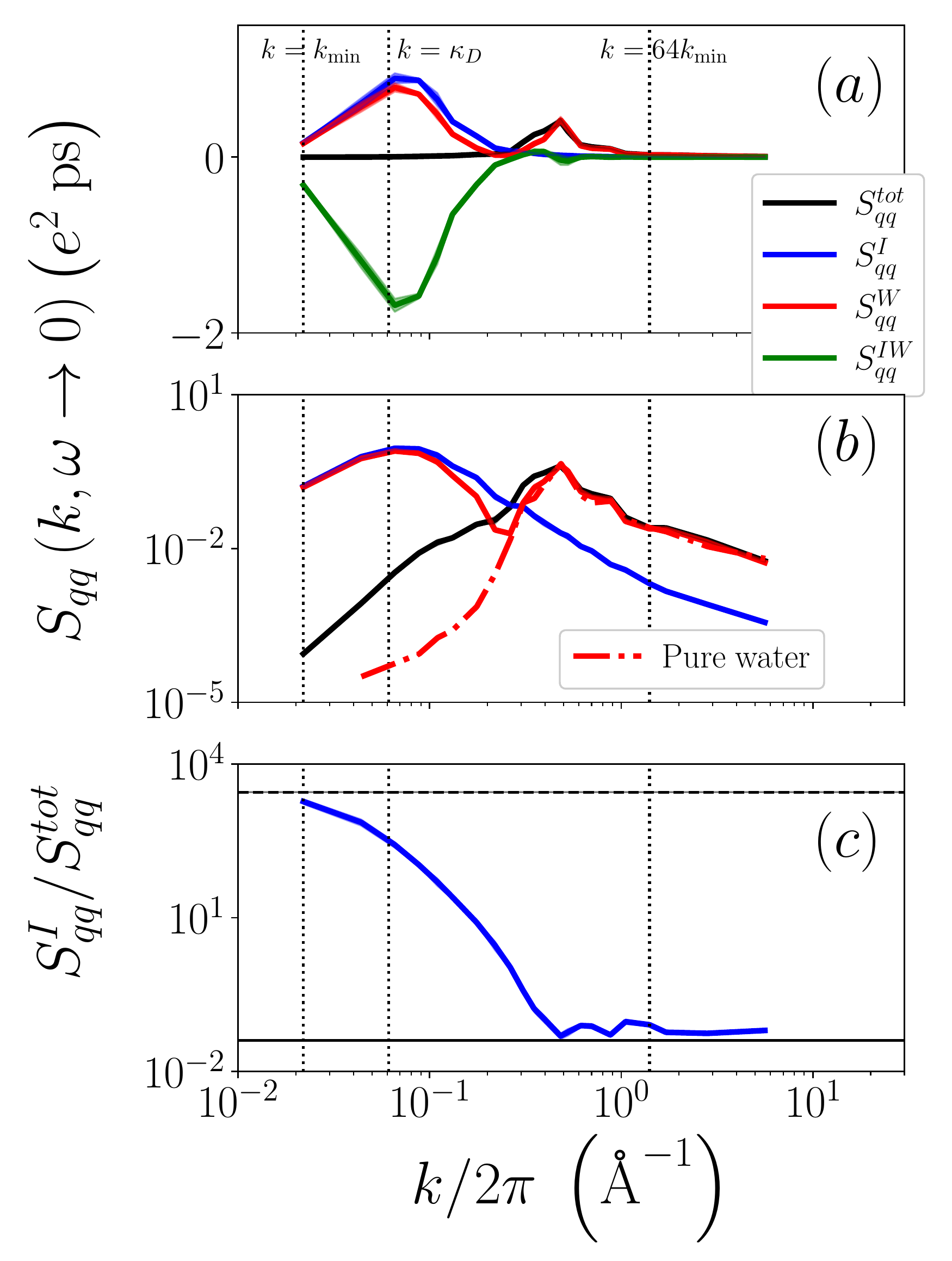}
    \caption{
    Zero-frequency limit of the dynamic charge-charge structure factor $S_{qq}(k,\omega\to0)$ for the MD simulations with an explicit solvent as a function of $k$ in log-lin scale (a) and log-log scale (b). Both panels show the total $S_{qq}^{tot}$ in black, as well as the contributions of ion-ion correlations, $S^{I}_{qq}$, in blue and water-water correlations, $S^{W}_{qq}$, in red (see Eq.~\ref{eq:SzzIandW}). The cross ion-water correlations, $S^{IW}_{qq}$, are also reported in green in panel (a), while panel (b) further shows the result for pure water (red dashed-dotted line). Panel (c) shows the ratio between the contribution of ions and the total, \emph{i.e.}
    $S_{qq}^{I}(k,\omega\to0)/S_{qq}^{tot}(k,\omega\to0)$.
    The vertical dotted lines in all panels indicate $k=k_{min}$ and $k=64k_{min}$ (considered in Fig.~\ref{fig:SzzContributions}), as well as $k=\kappa_D$, the inverse Debye screening length. The top and bottom horizontal lines in panel (c) indicate the the square of the static permittivity $\epsilon_s^2$ using the value determined from the simulations at finite concentration (Eq.~\ref{eq:CalcEpsilon}), as well as the ratio $(\sum_{i\in I} q_i^2)/(\sum_{i\in all} q_i^2)$, respectively.
    }
    \label{fig:SzzContributionsZeroFreq}
\end{figure}

Fig.~\ref{fig:SzzContributions}a and~\ref{fig:SzzContributions}b report the total $S_{qq}(k,\omega)$ and the ion-ion and water-water contributions for $k=k_{min}$ and $k=64k_{min}$, which are representative of the $k\to0$ and $k\to\infty$ regimes, respectively. The cross-correlations, which are generally negative (see below) are not shown on this log-log scale, but can be inferred from the other terms. For both the small and large $k$ regimes, at high frequency the total charge fluctuations correspond essentially to that of water only, and the ion-ion and cross terms are negligible. In fact, this observation holds for the whole frequency range at large $k$ (see also the discussion of Fig.~\ref{fig:SzzContributionsZeroFreq}c below for the $\omega\to0$ limit). This is not the case for $k\to0$: at low frequency, the ion-ion and water-water contributions are similar and much larger than the total $S_{qq}(k,\omega)$, which points to the importance of the cross term.

The significance of ion-water correlations is further illustrated in Fig.~\ref{fig:SzzContributionsZeroFreq}a, which reports all contributions (as well as the total) to the low frequency limit $S_{qq}(k,\omega\to0)$ as a function of $k$. For $k/2\pi\lesssim 0.3$~\AA$^{-1}$, the cross term is negative and almost compensates the sum of the other two contributions, which are comparable ($S^{tot}_{qq}\ll S^{W}_{qq}\approx S^{I}_{qq}$). In the static limit, ion-ion interactions are screened by the dielectric solvent and water is also dramatically impacted by the presence of ions, both in their immediate vicinity with the formation of solvation shells and at longer distances due to the screening of the electric field by the ions, which modifies the dipolar (and higher-order) correlations between solvent molecules. This static mutual screening can be captured with liquid state theories, \emph{e.g.} using integral equations~\cite{belloni_screened_2018, borgis_what_2018, simonin_solution_2020, simonin_full_2021}, or in the ``dressed ion'' picture of Kjellander, who also emphasized the key role of non-local electrostatics~\cite{kjellander_exact_1992, kjellander_fundamental_2007, kjellander_nonlocal_2016, kjellander_focus_2018, kjellander_intimate_2019, kjellander_multiple_2020}. 

The zero-frequency limit $S_{qq}(k,\omega\to0)$ is not the static charge-charge structure factor, which corresponds to $t=0$ rather than $\omega=0$, but reflects the mutual screening of ions and water in the presence of a static external electric field. In the $(k,\omega)\to0$ limit, these correlations between the fluctuations of the polarization due to water dipoles and ionic displacements ($\vec{P}_W$ and the itinerant polarization $\vec{P}_I$, respectively) are reflected in the Stillinger-Lovett conditions~\cite{stillinger_ionpair_1968,stillinger_general_1968}. Following Refs.~\cite{caillol_theoretical_1986,cox_finite_2019}, these sum rules can be expressed as $\langle \vec{P}_{W}\cdot \vec{P}_{I} \rangle= - \langle | \vec{P}_{W}|^2 \rangle$ and $\langle | \vec{P}_{W}|^2 \rangle = \langle |\vec{P}_{I}|^2 \rangle - 3\epsilon_0 k_BT/V$. The results displayed in Fig.~\ref{fig:SzzContributionsZeroFreq}a show that these (anti-)correlations between water and ions persist at finite $k$ corresponding to distances larger than the molecular sizes. Note that in our analysis $S^{IW}_{qq}$ includes ion-water and water-ion terms, which explains the factor of two with respect to the first sum rule.
 
Fig.~\ref{fig:SzzContributionsZeroFreq}b further shows the same results as panel~\ref{fig:SzzContributionsZeroFreq}a (except the cross term) in log-log scale, and compares them to the results for pure water. While the behaviour of water in the ionic solution is similar to that of pure water for large $k$ (probing length scales similar or smaller than the intramolecular distances), its contribution to $S_{qq}(k,\omega\to0)$ follows that of the ions for $k\to0$ (typically $k\lesssim\kappa_D$). This again illustrates the above-discussed correlations between water and ions fluctuations. In this small $k$ regime, $S^{I}_{qq}\approx S^{W}_{qq}$, and the total $S^{tot}_{qq}$, while much smaller, displays a similar decay with $k$. As seen in Section~\ref{sec:Results:ExplicitImplicit}, $S^{I}_{qq}(k,\omega)$ is reasonably well described in this regime by the implicit solvent model based on the static permittivity $\epsilon_r$ of the pure solvent. While this is clearly not sufficient to describe the charge fluctuations for larger $k$, one may examine whether the total charge fluctuations in the limit $k\to0$ and $\omega\to0$ can be expressed from the sole contribution of the ions. For example, in the static limit the electrostatic potential or the field due to a point ion in a solvent can be expressed as that of the bare ion divided by the permittivity, because the contribution of the solvent is $(-1+1/\epsilon_r)$ times that of the ion (see also Eq.~\ref{eq:ChiL}) -- and almost cancels the latter for $\epsilon_r\gg1$ as in the case of water. 

Fig.~\ref{fig:SzzContributionsZeroFreq}c reports the ratio between the contribution of ions and the total charge-charge dynamic structure factor in the $\omega\to0$ limit, \emph{i.e.} $S_{qq}^{I}(k,\omega\to0)/S_{qq}^{tot}(k,\omega\to0)$. For $k\to\infty$, this ratio converges to a plateau, consistent with the value expected by assuming that in this limit of infinitely short length scales only the ``self'' term for each atom contributes to the product defining the charge-charge structure factor (see Eq.~\ref{eq:DefFzz}), which results in a ratio $(\sum_{i\in I} q_i^2)/(\sum_{i\in all} q_i^2)\approx 0.04$. More interesting is the opposite limit $k\to0$, where the ratio  $S_{qq}^{I}/S_{qq}^{tot}$ seems to reach a plateau. Since the ionic contribution is qualitatively well described in this limit by the PNP result (see Fig.~\ref{fig:SzzComparison}b), one can use Eq.~\ref{eq:SzzPNP} to estimate the zero-frequency limit, which scales as $\kappa_D^{-4}\propto\epsilon_r^2$. One can therefore conjecture that this factor also corresponds to the ratio between the bare contribution of the ions (in vacuum) and that of the ions in solution, as explained above for the screened potential. Fig.~\ref{fig:SzzContributionsZeroFreq}c also shows the plateau corresponding to $\epsilon_s^2$, with the permittivity of the ionic solution $\epsilon_s=53.0\pm0.3$ (obtained by Eq.~\ref{eq:CalcEpsilon}), smaller than that of the pure solvent by a factor consistent with previously reported results at this concentration~\cite{kalcher_structure-thermodynamics_2009,sega_calculation_2013}. Even though the range of $k$ is limited by the finite size of the simulation box and the logarithmic scale does not allow to appreciate the exact value of the plateau, the consistency with the numerical results supports the above discussion. We note that the latter neglects the $k$-dependence of the static permittivity, \emph{i.e.} non-local electrostatic effects whose importance was highlighted by several authors~\cite{bopp_static_1996, kjellander_decay_2016, berthoumieux_dielectric_2019, vatin_electrostatic_2021}. The possibility to analyze the contributions to $S_{qq}^{tot}(k,\omega)$ using molecular simulations can shed light on how to improve continuum descriptions not only of the static permittivity, but also on its dynamic response.

\section{Conclusions}

We have illustrated the role played by electric fluctuations in a number of experiments, which probe various observables that all reflect the same underlying dynamics of ions and solvent molecules. The microscopic fluctuations of the charge are encoded in the charge-charge intermediate scattering function or the charge-charge dynamic structure factor, $S_{qq}(k,\omega)$. While these quantities cannot be measured directly as a function of the wavenumber and time or frequency, many observables can be expressed as special cases ($k\to0$ for the macroscopic limit, $\omega\to0$ for the static limit) or as integrals over modes that depend on the property of interest. In this work, we illustrated this on a few examples to highlight the relevance of combining seemingly unrelated experiments that provide complementary windows on the microscopic charge fluctuations.

We discussed several theoretical approaches to model the dynamics of charge fluctuations in electrolytes, and presented new simulation results with both explicit and implicit solvent models for a $\approx1$~M aqueous NaCl. As expected for this rather high concentration, the linearized Poisson-Nernst-Planck theory cannot predict quantitatively the charge-charge dynamic structure factor over the whole wavenumber and frequency range. Nevertheless, it captures the main features for the small $k$ and $\omega$ regimes. The predictions for the $S_{qq}(k,\omega)$ in the intermediate $k$ range can be significantly improved by introducing the static correlations obtained from Langevin or Brownian dynamics in the more general result of Dynamic Density Functional Theory Eq.~\ref{eq:SzzDDFTapprox}. This suggests that the main limitation of linearized PNP is the corresponding free energy functional and not the description of dynamics itself in the considered case. However, the implicit-solvent simulations neglect other important features related to the solvation of ions by the molecular solvent (at large $k$ and $\omega$) and hydrodynamic couplings between ions via the solvent (at small $k$ and $\omega$). This is reflected in the fact that introducing the static correlations from molecular simulations in the DDFT result is not sufficient to quantitatively predict the dynamic structure factor.

Finally, we analyzed with molecular dynamics simulations the contributions of ion-ion, water-water and ion-water correlations to the total charge-charge dynamic structure factor. Even at this relatively high concentration, $S_{qq}(k,\omega)$ is dominated by water for all frequencies for large $k$, as well as for high frequency at all $k$. In contrast, for small $k$ and $\omega$ the total $S_{qq}(k,\omega)$ is much smaller than both the ion and water contributions, which are comparable, due to the strong negative correlation between them. We discussed these results in the general context of screening, such as exact sum rules (Stillinger-Lovett conditions) in the $(k,\omega)\to0$ limit arising from the mutual influence of ions and water. These correlations are here shown to persist for finite wavenumbers corresponding to distances larger than the molecular sizes. They further suggest that in the $(k,\omega)\to0$ limit, it remains possible to relate the total $S_{qq}(k,\omega)$ to the ion contribution only, with a scaling factor involving the static permittivity, thereby making the link with the PNP-like description. The possibility to analyze the contributions to $S_{qq}^{tot}(k,\omega)$ using molecular simulations can shed light on how to improve continuous descriptions not only of the static permittivity, but also on its dynamic response. 

The examples developed in the present work are mainly related to dielectric spectroscopy and impedance measurements, even though NMR relaxation provides an illustration of a very different type of experiments. They are but a few of the many possibilities mentioned to obtain information on the microscopic dynamics of charges in ionic fluids -- which extend well beyond the important case of aqueous electrolytes. In particular, the cross-correlations between charge and other properties such as mass or momentum can be probed in electrokinetic/electroacoustic experiments, and are related to the electrostatic contribution to the friction exerted on the ions~\cite{sedlmeier_charge/mass_2014, samanta_ionic_2022}, while specific information can be obtained using other experiments such as quasi-elastic neutron scattering, to which hydrogen atoms contribute significantly. We hope that the present transverse perspective on the dynamics in ionic fluids will motivate experts of different experimental techniques to combine their complementary views on the same systems.


\section*{Author Contributions}

\textbf{Th\^e Hoang Ngoc Minh:} Conceptualization (equal); Formal Analysis (equal); Investigation (lead); Methodology (equal); Validation (equal); Writing/Original Draft Preparation (equal); Writing/Review \& Editing (supporting);
\textbf{Jeongmin Kim:} Conceptualization (equal); Formal Analysis (equal); Investigation (lead); Methodology (equal); Validation (equal); Writing/Original Draft Preparation (equal); Writing/Review \& Editing (supporting);
\textbf{Giovanni Pireddu:} Conceptualization (equal); Validation (equal); Writing/Original Draft Preparation (supporting); Writing/Review \& Editing (supporting);
\textbf{Iurii Chubak:} Conceptualization (equal); Validation (equal); Writing/Original Draft Preparation (supporting); Writing/Review \& Editing (supporting);
\textbf{Swetha Nair:} Validation (supporting); 
\textbf{Benjamin Rotenberg:} Conceptualization (lead); Formal Analysis (equal); Funding Acquisition (lead); Investigation (supporting); Methodology (equal); Supervision (lead); Validation (equal); Writing/Original Draft Preparation (lead); Writing/Review \& Editing (lead).

\section*{Conflicts of interest}
There are no conflicts to declare.

\section*{Acknowledgements}
The authors thank Sophie Marbach, Pierre Illien, Antoine Carof, Lyd\'eric Bocquet and Susan Perkin for useful discussions. This project received funding from the European Research Council under the European Union’s Horizon 2020 research and innovation program (project SENSES, grant Agreement No. 863473). The authors acknowledge access to HPC resources from GENCI-IDRIS (grant no. 2022-AD010912966R1).

\balance


\bibliographystyle{aipnum4-1} %

\end{document}